\documentclass[namedreferences]{solarphysics}

\usepackage[hyperref,optionalrh,natbib]{spr-sola-addons} 
\usepackage{graphicx}        
\usepackage{epstopdf}
\usepackage{color}           
\usepackage{ulem}

\newcommand{\etal}{{\it et al.}}

\newcommand{\apj}{    {\it Astrophys. J.}}

\newcommand{\pasj}{   {\it Pub. Astron. Soc. Japan}}

\newcommand{\solphys}{{\it Solar Phys.}}

\newcommand{\ssr}{    {\it Space Sci. Rev.}}

\ifx \doiurl    \undefined \def
\doiurl#1{\href{http://dx.doi.org/#1}{\textsf{DOI}}}\fi \ifx
\adsurl    \undefined \def
\adsurl#1{\href{http://adsabs.harvard.edu/abs/#1}{\textsf{ADS}}}\fi
\ifx \arxivurl  \undefined \def
\arxivurl#1{\href{http://arxiv.org/abs/#1}{\textsf{arXiv}}}\fi

\begin{document}

\begin{article}

\begin{opening}

\title{Flare SOL2012-07-06: on the origin of the circular polarization reversal between 17 GHz and 34 GHz
  \\ }

\author{A.~\surname{Altyntsev}$^{1}$\sep
        N.~\surname{Meshalkina}$^{1}$\sep
        I.~\surname{Myshyakov}$^{1}$\sep
        V.~\surname{Pal`shin}$^{2}$\sep
        G.~\surname{Fleishman}$^{3,2}$\sep
        }
\runningauthor{A. Altyntsev et al.} \runningtitle{Flare
SOL2012-07-06}

   \institute{$^{1}$ Institute of Solar-Terrestrial Physics SB RAS, Lermontov st.\ 126A, Irkutsk 664033, Russian Federation
                     email: \url{altyntsev@iszf.irk.ru} email: \url{nata@iszf.irk.ru} email: \url{ivan_m@iszf.irk.ru} \\
                        $^{2}$ Ioffe Physical Technical Institute, St. Petersburg, 194021, Russian Federation
                     email: \url{e.mail-a} email: \url{val@mail.ioffe.ru}\\
                     $^{3}$ Center For Solar-Terrestrial Research, New Jersey Institute of Technology, Newark, NJ
                     07102\\
                     email: \url{e.mail-a} email: \url{gfleishm@njit.edu}\\
             }

\begin{abstract}
The new generations of multiwavelength radioheliographs with high
spatial resolution will employ microwave imaging spectropolarimetry
to recover flare topology and plasma parameters in the flare sources
and along the wave propagation paths. The recorded polarization
depends on the emission mechanism and emission regime (optically
thick or thin), the emitting particle properties, and propagation
effects. Here, we report an unusual flare, SOL2012-07-06T01:37,
whose optically thin gyrosynchrotron emission of the main source
displays an apparently ordinary mode sense of polarization in
contrast to the classical theory that favors the extraordinary mode.
This flare produced copious nonthermal emission in hard X-rays and
in high-frequency microwaves up to 80 GHz. It is found that the main
flare source corresponds to an interaction site of two loops with
greatly different sizes. The flare occurred in the central part of
the solar disk, which allows reconstructing the magnetic field in
the flare region using vector magnetogram data.
 We have investigated the three possible known reasons of
 the circular polarization sense reversal --  mode coupling,
  positron contribution, and the effect of beamed angular distribution.
We excluded polarization reversal due to contribution of positrons
because there was no relevant response in the X-ray emission.  We
find that a beam-like electron distribution can produce the observed
polarization behavior, but the source thermal density must be much
higher than the estimate from to the X-ray data. We conclude that
the apparent ordinary wave emission in the optically thin mode is
due to radio wave propagation across the quasi-transverse (QT)
layer. The abnormally high transition frequency (above 35 GHz) can
be achieved reasonably low in the corona  where the magnetic field
value is high and transverse to the line of sight. This places the
microwave source below this QT layer, \textit{i.e.} very low in the
corona.

\end{abstract}
\end{opening}

\section{Introduction}
     \label{S-Introduction}

The generation of high-energy charged particles in solar flares
remains an outstanding, high-priority problem of solar physics. Many
competing acceleration mechanisms have been proposed and compared
with various data sets, but no ultimate choice has been made in
favor of one or only a few mechanisms. The flare microwave emission
of mildly relativistic electrons can help to distinguish these
acceleration mechanisms as it  provides us with important
information on plasma and nonthermal particles in the flare region,
and coronal plasma properties along the wave path.

The spatially resolved data of the Stokes $I$ and $V$ components are
available from a number of solar radio instruments at different
microwave frequencies, such as the \textit{Nobeyama Radioheliograph}
(NoRH,
 \opencite{Nakajima94}), the \textit{Siberian Solar Radio
 Telescope}
(SSRT, \opencite{Grechnev03}), and \textit{Owens Valley Solar
Arrays} (OVSA, \opencite{Gary94}). A new generation of microwave
multi`wavelength radioheliographs is now under various stages of
development: the expanded OVSA (\opencite{Gary12}), the
\textit{Frequency Agile Solar Radiotelescope} (FASR,
\opencite{Bastian04}), the \textit{Mingantu Ultrawide Spectral
Radioheliograph} (MUSER, \opencite{Yan13}) and the \textit{Siberian
Radio Heliograph} (SRH, \citeauthor{Lesovoi14}, \citeyear{Lesovoi14,
Lesovoi17}).

Gyrosynchrotron sources at high frequencies are often located close
to the flare loop footpoints where the magnetic field exceeds
several hundred Gauss, before lower frequencies are emitted from
greater heights. Flares initiated by an interaction of two loops of
rather different size have been reported (\opencite{Hanaoka97};
\opencite{Nishio97}; \opencite{Altyntsev16};
\opencite{Fleishman16a}). The NoRH polarization images at 17 GHz
showed footpoints of a small loop and the SSRT observed a large loop
at 5.7 GHz. The main microwave and X-ray sources were located in  or
close to the site of the loop--loop interaction.

The measured polarization of the microwave sources depends on
intrinsic polarization at the emission source and its modification
through propagation effects. The intrinsic polarization is
determined by the emission mechanism and emission regime
(\textit{e.g.}, optically thin or thick regimes), the emission
'driver' properties, \textit{e.g.}, the type of radiating particles
(electrons or positrons) or their angular distribution, and the
source properties, \textit{e.g.}, the strength and orientation of
the magnetic field. It is known that the continuum microwave
emission of solar flares is primarily generated by the nonthermal
electrons through the gyrosynchrotron emission mechanism
(\opencite{Ramaty}). The gyrosynchrotron emissivity increases with
the magnitude of the magnetic field. A compact microwave
high--frequency source is therefore expected to appear at the loop
footpoint, where the nonthermal particles precipitate and the
magnetic field is enhanced. Emission from these sources is
circularly polarized, and the sign of the polarization is determined
by the direction of the magnetic field at the source.

In the case of an isotropic distribution of emitting electrons the
gyrosynchrotron mechanism favors the extraordinary mode in the
optically thin regime at high frequencies.  Thus, the source located
at the footpoint of a loop rooted in the north (positive) polarity
magnetic field should be seen as the right--hand circular
polarization  (RCP) source (\opencite{Dulk82};
\opencite{Bastian98}). In some cases, however, the microwave source
can favor the ordinary waves. First of all, the optically thick
polarization at frequencies below the spectrum peak is polarized in
the sense of the ordinary mode. In the optically thin regime, the
sense of polarization can correspond to the ordinary wave mode  for
a certain range of viewing angles for beamlike distributions of
emitting nonthermal electrons (\opencite{Fleishman03}). Such an
event with a polarization reversal at high frequencies, which
corresponds to ordinary--mode emission, was found by
\cite{Altyntsev08} and recently by \cite{Morgachev14} and the
reversal was explained by the strong anisotropy in the electron
pitch--angle distribution. Then, \cite{Fleishman13} proposed that
waves with the sense of the ordinary wave can be generated at high
frequencies by relativistic positrons if their differential energy
spectrum at high energies exceeds that of the electrons.

The importance of propagation effects was discovered a long time ago
(\opencite{Cohen}; \opencite{Zheleznyakov};
\opencite{Alissandrakis93}). A circularly polarized wave reverts its
helicity when it propagates through a quasi--transverse region,
where the directions of ray propagation and of the magnetic field
are orthogonal. If the mode coupling is weak the polarization
continuously adjusts itself to the local polarization state for the
mode in question. If the mode coupling is strong, there is no change
in the state of polarization of radiation through propagation
effects. \cite{Cohen} defined the transition frequency $f_{t}$ such
that for $f \gg f_{t}$ ($f \ll f_{t}$) the coupling is strong
(weak). The polarization reversals are used to probe the corona
(see, \textit{e.g.} \opencite{Lee98}; \opencite{Ryabov05}).

The NoRH images at 17 GHz of the Stokes $I$ and $V$ components are
currently the most suitable images for studying polarization effects
because this frequency typically corresponds  to the optically thin
regime and is expected to be not strongly affected by the mode
coupling. The NoRH images have high spatial resolution and have been
intensively studied by solar astronomers for more than twenty years.
Until recently, only very few studies paid particular attention to
the radio polarization observed by the NoRH, however. \cite{Huang09}
published a statistical analysis of the flaring loops. As mentioned
above it is theoretically predicted that the gyrosynchrotron
emission is dominated by the extraordinary mode. The footpoints of
loops should be located on the south (negative) and north (positive)
magnetic field and the emission of the conjugate points must have
the opposite sense of circular polarization. Most of the analyzed
events had the same polarization sense at both footpoints, however.
\cite{Huang09} did not study this effect in detail, but have
emphasized that the polarization of the microwave burst was strongly
affected by propagation effects through the overlying magnetic
field, such as linear mode conversion.

Thus the study of the polarization and its reversals is a
potentially powerful, but yet largely unexplored, diagnostic tool.
Of particular interest is the study of events recorded by the
\textit{Nobeyama Radio Polarimeters} (NoRP, \opencite{Torii79})
with opposite senses of circular polarization at 17 and
 35 GHz. Fleishman et al. (2013) have recently searched
  the entire Nobeyama database until 2013 and have found
   that such a reversal was observed for about 1\% of all
    microwave bursts recorded by the Nobeyama instruments.

This article analyzes the M2.9 impulsive flare on 2012 July 6 at
01:38 UT, in which the microwave burst was oppositely circularly
polarized at 17 and 35 GHz. The flare sources in X--rays and
microwaves were separated by a distance sufficient to resolve them
in the RHESSI and Nobeyama images. The event occurred reasonably
close to the disk center to allow reconstructing the magnetic field
in the flare region. In the flash (impulsive) phase of the flare a
significant burst was observed in microwaves up to 80~GHz, and in
X--rays in the range of up to 300--800 keV.

\section{Instruments and observations}

Microwave total fluxes were recorded with the \textit{Nobeyama Radio
Polarimeters} in intensity $I=R+L$ and circular polarization $V=R-L$
at six frequencies (1, 2, 3.75, 9.4, 17, and 35~GHz) and in the
intensity
 only at 80~GHz with a time resolution of 0.1~sec, and in
 1~s intensity data from the \textit{Radio Solar Telescope
 Network} (RSTN, \opencite{Guidice81}) at seven
 frequencies (0.4, 0.6, 1.4, 2.7, 5.0,8.8, and 15.4~GHz).
 To measure the spectrum, we also used data from
 the \textit{Solar Radio Spectropolarimeters} (SRS)  ($I$ and $V$ fluxes
 at 16 frequencies: 2.3, 2.6, 2.8, 3.2, 3.6, 4.2, 4.8, 5.6, 6.6, 7.8, 8.7,
 10.1, 13.2, 15.7, 19.9, and 22.9 GHz;
 the temporal resolution is 1.6 s) that have been created by the SSRT team.

The microwave imaging was performed with the SSRT
(\opencite{Grechnev03}) at 5.7~GHz (intensity and polarization) and
the NoRH (\opencite{Nakajima94}) at 17 GHz (intensity and
polarization) and 34~GHz (intensity only). The NoRH spatial
resolution was  13 arcsec at 17 GHz and 8 arcsec at 34 GHz. We used
the imaging with 1 s cadence. The 17 GHz fluxes in intensity and
polarization obtained from the NoRH images are close to the NoRP
data. The 35 GHz intensity flux measured by the NoRP is 1.9 times
higher than the flux calculated from the NoRH images at 34 GHz,
however. We have found that the NoRH disk--fitting was sufficiently
reliable before the burst maximum. At the maximum, the partial
images showed the reliable temperature of the quiet Sun as well. We
therefore reduced the NoRP fluxes at 35 GHz in intensity and
polarization using this factor. This correction does not affect the
values of the polarization degree.

The SSRT is a cross--shaped interferometer and the data recorded
 by the east--west (EW) and north--south (NS) arrays provide two--dimensional
  images of the solar disk every two to three minutes and one--dimensional
   images
(scans) every 0.3~s in the standard mode of the observations.
The receiver system of SSRT contains a spectrum analyzer with 250
 frequency channels, which corresponds to the knife--edge--shaped
 fan beams for the NS and EW arrays. The response at each frequency
  corresponds to the emission from a narrow strip on the solar disk
   whose position and width depend on the observation time, array type,
    and frequency. The signals from all the channels are recorded
    simultaneously and generate a one--dimensional distribution of
    solar radio brightness. In this study we use the one--dimenshional images
    provided by the EW linear antenna array. The width of
    the EW beam was 26 arcsec and the receiving strip was made at an
    angle of 28 degrees counterclockwise relative to the central solar meridian.

We used the HXR data obtained with the RHESSI (\opencite{Lin02}).
For imaging we used the forward--fit (FWD) method and for spectrum
calculations -- we used the OSPEX package. The FWD approach is
rather effective for sources with a relatively simple structure such
as the one we study here. HXR and gamma--ray data for this event
were also obtained with the \textit{WIND}/KONUS spectrometer
(\opencite{Aptekar95}; \opencite{Palshin14}).

To study structure of the flare region we use the EUV images
obtained\ with the \textit{Atmospheric Imaging Assembly} (AIA,
\opencite{Lemen}) onboard the \textit{Solar Dynamics Observatory}
(SDO). For the boundary conditions, the 12--minute full--Sun vector
magnetograms obtained by the SDO/HMI instrument
(\opencite{Scherrer}) were used.

\begin{figure}    

 \centerline{\includegraphics[width=\textwidth]{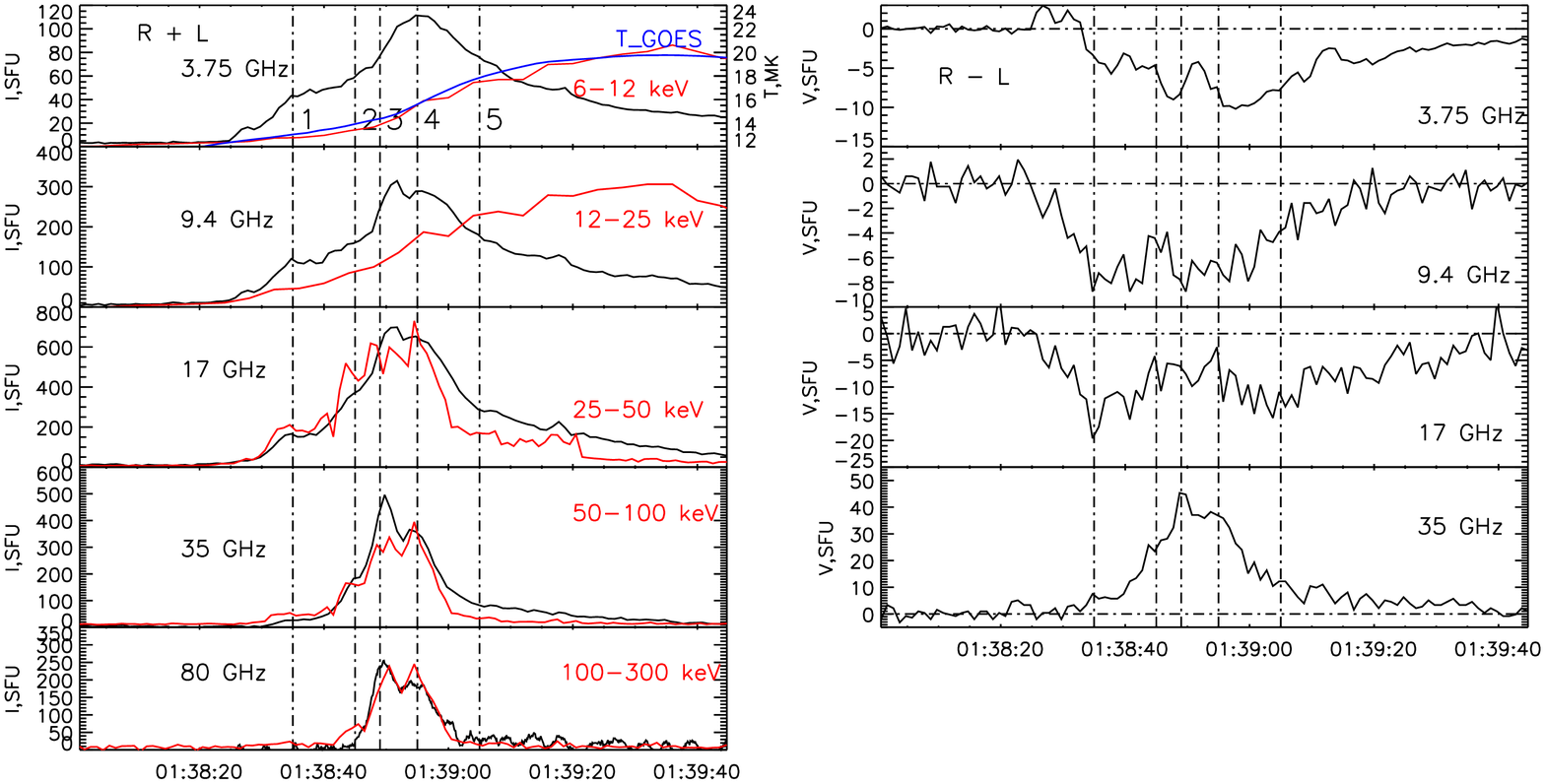}
             }
             \caption{Lightcurves of X--ray (RHESSI, \textit{red}) and microwave fluxes (NoRP, \textit{black}) in intensity ($R + L$, \textit{left})
             and in polarization ($R - L$, \textit{right}).
             Plasma temperature profile calculated from the GOES data are shown in the left top panel.
             The RHESSI data are presented in arbitrary units.
                     }
  \label{F1--simple}
  \end{figure}

Light curves recorded in the flash stage with the NoRH and the
RHESSI are presented in Figure~\ref{F1--simple}. The duration and
behavior of the HXR signals were strongly different at energies
above and below 25~keV. The shortest signals of 10--15 s duration
were observed at high energies. At low energies the flux started to
grow smoothly before and reached the maximum a minute after the
impulsive peak. Note that the behavior of the hard X--rays with
energies lower than 12 keV is similar to that of the plasma
temperature provided by the GOES data. The temperature reached 20 MK
at 01:39:37 UT.

Microwave emission was observed in a wide range of frequencies up to
80 GHz, which is the highest working frequency of the Nobeyama
spectropolarimeters. The light curves at high frequencies were
similar to the hardest X--ray ones. The maximum radio flux was
observed at 17 GHz and reached 730 SFU at 01:38:50 UT. The duration
of the high--frequency burst at 35 -- 80 GHz was the same as that of
the hard X--ray burst. At frequencies below 17 GHz, the microwave
light curves became smoother with a relatively slower rise and
decay. The polarization profiles at frequencies below 35 GHz were
left--handed during the impulsive phase. The polarization degree did
not exceed a few percent. At 35 GHz, the polarization profile was
right--handed and showed a short pulse similar to the intensity
pulse, and the degree of polarization was $V/I = 0.1 - 0.35$. The 80
GHz data do not have polarization information.

Spectral properties. The ratio of the RHESSI signals at 100--300 keV
and 50--100 keV shows a soft--hard--soft behavior. Using the OSPEX
procedure we found that the hard X--ray photons with energies up to
600 keV can be described as $2.22\times(E/50)^{-3.0}$
ph/(s$\cdot$cm$^{2}$keV) at flux maximum. The \textit{Wind}/KONUS
spectrum $2.15\times(E/50_{keV})^{-3.2}$ ph/(s$\cdot$cm$^{2}$keV) is
consistent with the spectrum of RHESSI. We searched for an
annihilation line and found that the light curves recorded by both
instruments do not show any noticeable enhancement above the
background at 511 keV during the flash phase.

\begin{figure}    

\centerline{\includegraphics[width=\textwidth]{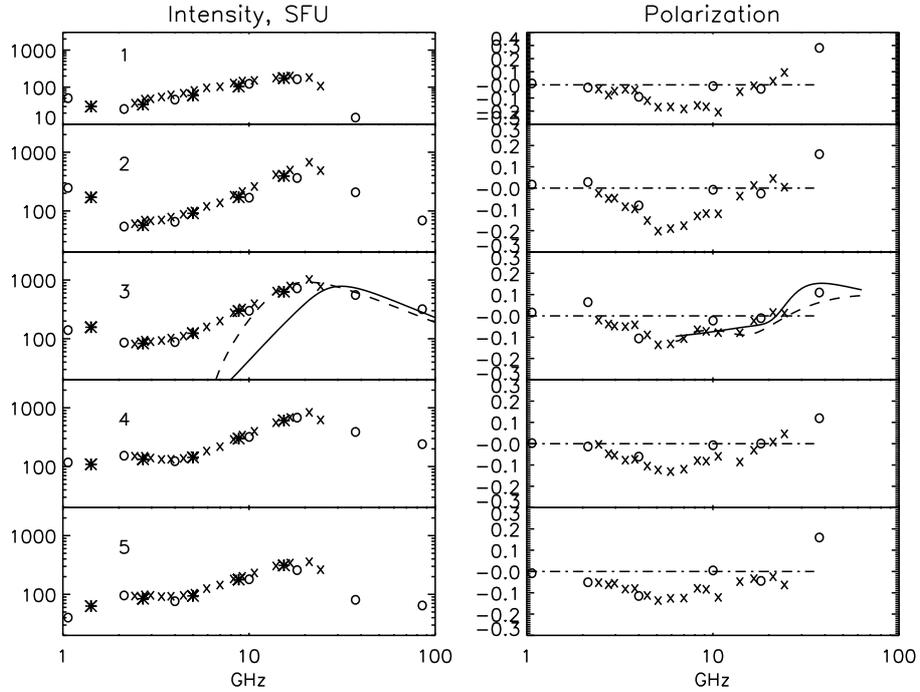}
             }
             \caption{Microwave spectra in intensity (left)
              and polarization degree (right) for the time
              frames marked in  Figure~\ref{F1--simple}. The data were recorded by the RSTN (asterisks),
              NoRP (circles) and the SRS spectropolarimeters (crosses). In panel 3 the results of
              fitting the high--frequency part are shown by the solid and dashed curves for the isotropic and anisotropic case, respectively.
                     }
  \label{F7--simple}
  \end{figure}

The microwave spectra are shown in Figure~\ref{F7--simple}. Note the
high consistency of the spectral data from the different
spectrographs. The spectrum shapes are typical for gyrosynchrotron
emission. The spectral peak frequency was varying around 17 GHz and
was higher at time peaks 3 and 4. The emission at 35 and 80 GHz
clearly corresponds to the optically thin regime.

The spectra of the circular polarization degree for the same time
frames are shown in the right panels. Mainly left--handed emission
reaches the level of 0.1--0.2 at 5--7 GHz. At the frequencies above
17 GHz the polarization sense reversal was observed at both
spectropolarimeters: NoRP and SRS. At 35 GHz the polarization degree
varied from 0.25 down to 0.1. If the angular distribution of the
emitting electrons is isotropic, a behavior of the circular
polarization sense like this is expected for a source located above
the N--polarity region.

\begin{figure}    

\centerline{\includegraphics[width=\textwidth]{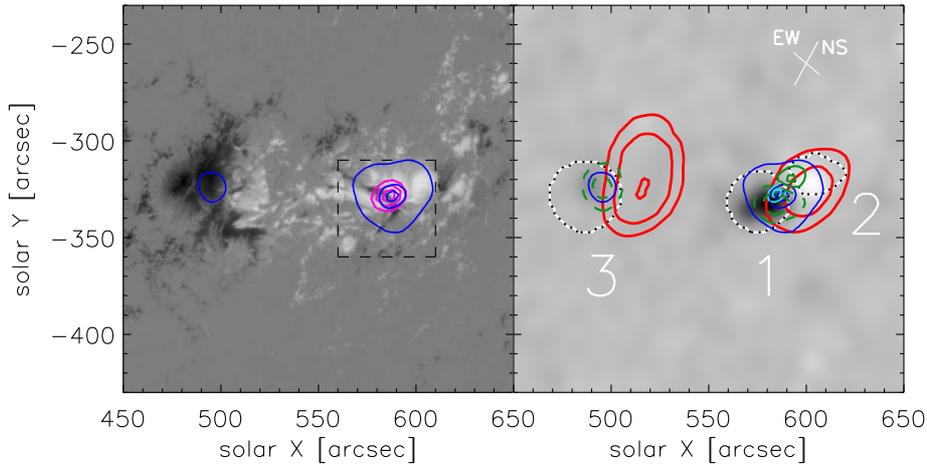}
             }
             \caption{Flare structure. The grayscale background is a magnetogram taken at 01:39:00 UT (left panel)
             and an RMS polarization map at 17 GHz (right). Blue contours mark
        the 17 GHz sources in intensity (levels $0.016,0.6,0.9)\times 120$ MK at 01:38:48 UT.
        In the left panel the pink contours present the HXR brightness
        levels at 25--50 keV (0.3, 0.6, 0.9, RHESSI) relative to maximum (averaged
        over 01:38:49 -- 01:38:54 UT).
        Right panel: the black--white contours bound the masks for flare sources,
             marked by numbers; red contours $(0.5,0.7,0.99)\times 142$ MK
        show the  intensity of microwave sources at 5.7 GHz
             (01:41 UT), and cyan contours $(0.6, 0.9)\times 9$ MK -- 34 GHz.
             Green contours present the 17 GHz emission in polarization at 01:38:48 UT:
        the dashed -- LCP, levels $(0.2,0.6)\times 3.4$ MK) and solid lines
        show RCP, levels $(0.3,0.9)\times 0.6$ MK). In the top right corner
        the cross presents directions and beam width of the one--dimensional SSRT scans.
        The extended frame (dashed black square in the left panel) is shown in  Figure~\ref{Frame--simple}.
                     }
  \label{F2--simple}
  \end{figure}

Flare configuration. The flare occurred in the very
flare--productive Active Region 11515 (S17W50). More than twenty
flares were observed on the day before the event under study
(including M--class flares). Microwave and X--ray sources aligned
with a cotemporal SDO/HMI magnetogram are shown at the peak time of
the burst in Figure~\ref{F2--simple}. The bulk of the emission was
observed from source 1, where the X--ray and 34 GHz sources and the
brightness centers of emission at 17 GHz were located. Microwave
source 1 was left--handedly polarized at 17 GHz. There were two more
flare sources, located to the northwest (source 2) at a distance of
15 arcsec and 90 arcsec eastward (source 3). Source 2 is detected in
polarization because of the opposite sense, \textit{i.e.} it was
right handed.

LCP source 3 occurred during the flare above an S--polarity region.
Figure~\ref{F2--simple} shows the contours of the first image at 5.7
GHz, which was not overexposed and was recorded at 01:41 UT. There
are two bright regions corresponding to  source 3 and the region
that combines  sources 1 and 2 because the spatial resolution was
too low to resolve them. The one--dimensional scans show that a
significant emission at 5.7 GHz was generated in the loops
connecting  source 3 with the region spanning sources  1 + 2.

\begin{figure}    

 \centerline{\includegraphics[width=\textwidth]{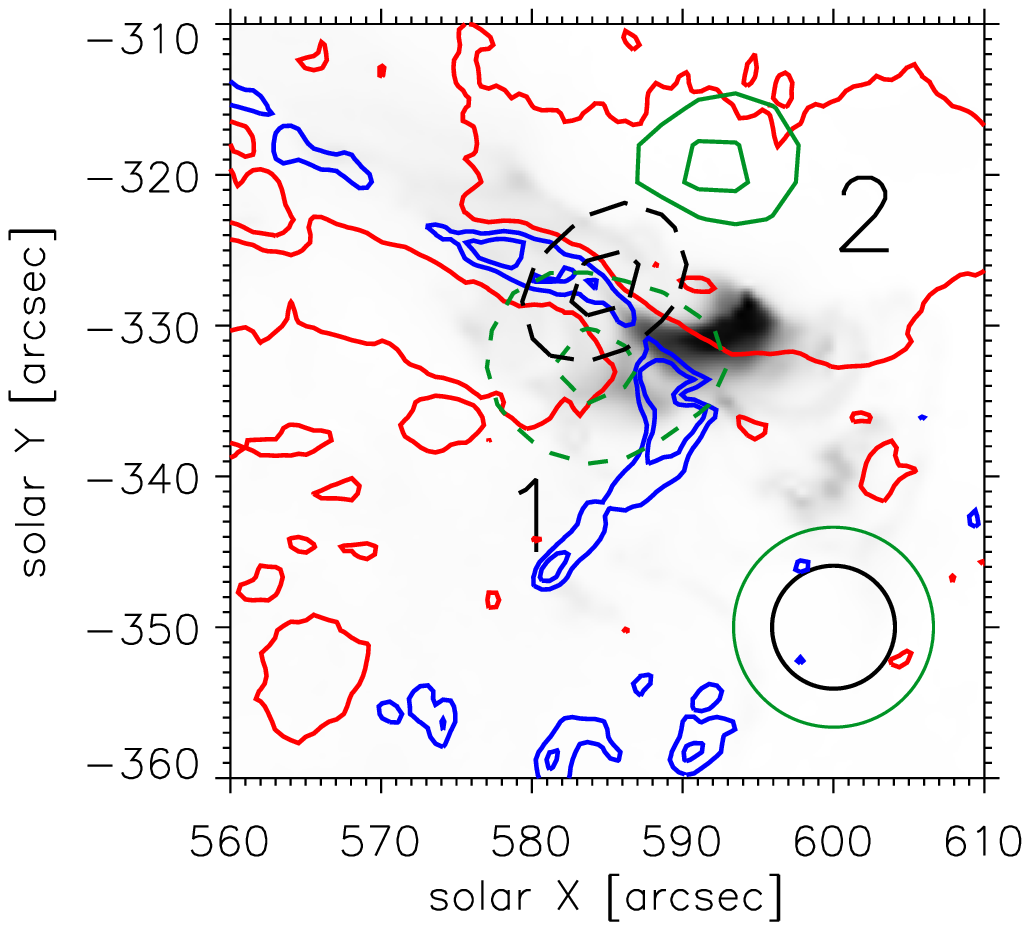}
             }
             \caption{ Flare sources 1 and 2. Background: AIA/SDO 335~\AA\ negative
image at 01:40:38 UT. The LOS magnetogram of HMI/SDO at 01:39:00 UT
is shown with blue contours at --500, --150 G and red contours at
500, 1500 G. The black dashed contours show the 34 GHz source at
$(0.5, 0.9)\times 46$ MK (01:38:48 UT). The brightness temperatures
at 17 GHz are shown for LCP $(0.5, 0.9)\times 3.4$ MK (green dashed
contours) and RCP, $(0.5, 0.9)\times 0.6$ MK (green solid contours).
The FWHM of the NoRH beams at  (green) and 34 GHz (yellow) are shown
in the right bottom corner.
                     }
  \label{Frame--simple}
  \end{figure}

The extended region with sources 1 and 2 is shown in
Figure~\ref{Frame--simple}. The background 335~\AA\ image
corresponds to a time frame in which the EUV images were not
saturated and the post--flare loop was clearly seen.  The EUV loop
connected a narrow S--polarity intrusion to a vast area of N
polarity. The brightness center of 34 GHz emission was located in
the S--polarity strip. The 17 GHz RCP source 2 was located in this
N--polarity region with a magnetic field $B
> 500$ G, the brightness center of the LCP source was in close
proximity to  the strip of S polarity. The distance between the
sources was about 15 arcsec. The apparent sizes of microwave sources
1 are close to the NoRH beam FWHM: 13 arcsec at 17 GHz and 8 arcsec
at 34 GHz. Thus the real sizes of the sources did not exceed a few
arcseconds.

\begin{figure}    

 \centerline{\includegraphics[width=\textwidth]{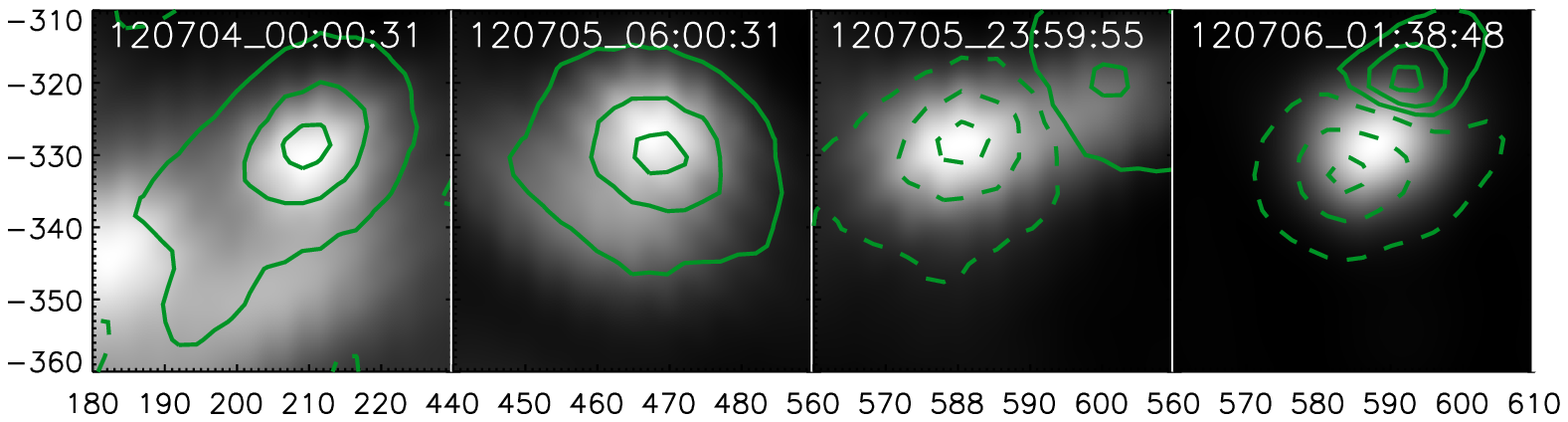}
             }
             \caption{ Evolution of sources 1 and 2 at 17 GHz.
             The grayscale background is the intensity. Contours correspond to RCP
              (solid) and LCP (dashed) emission. Levels 0.1, 0.5,
              and 0.9 are relative to the corresponding brightness maximum. Coordinates
are given in arcseconds from the solar center.
                     }
  \label{QTreversal}
  \end{figure}

A microwave source at 17 GHz in the place of source 1 appeared two
days before the flare and changed its polarization from RCP to LCP
on 5 July (Figure~\ref{QTreversal}). The source brightness
temperature was increasing from 80 000 K up to a few 100 000 K at
01:37:31 UT on July 6.

\begin{figure}    

 \centerline{\includegraphics[width=\textwidth]{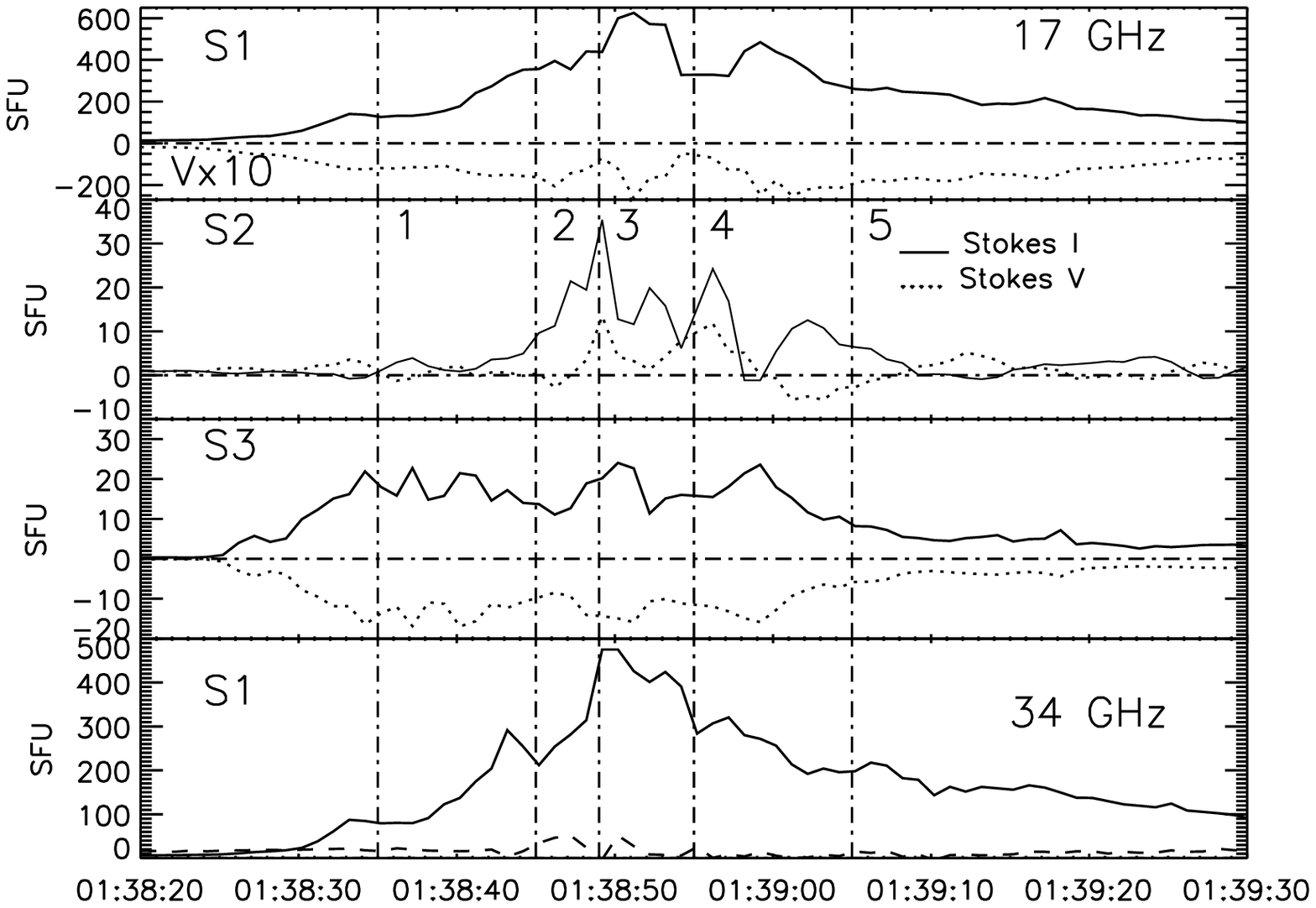}
             }
             \caption{Microwave fluxes from the flare sources marked in Figure~\ref{F2--simple}.
             In the bottom panel the solid curve corresponds to emission from source 1 bounded
             by the black--white contour in Figure~\ref{F2--simple}, and the dashed curve shows the
             profile of flux emitted outside the source.
                     }
  \label{F3--simple}
  \end{figure}

To reveal links between the sources during the flare we made a
so--called mean square (RMS) map using the 17 GHz images in
polarization (see, \textit{e.g.} \opencite{Grechnev03a}). The
sources are marked by the white--black dotted contours  that
corresponded to the half--width of the RMS brightness temperature
variances (Figure~\ref{F2--simple}). The time profiles for each
microwave source are shown in Figure~\ref{F3--simple}. There was an
impulsive component in the emission from source 2, and a gradual
component in the emission from source 3. The polarization degrees of
the emission from the two weak sources reached high values of 50\%
and more. The polarization degree of source 1 was about 5\% at 17
GHz. The brightening of source 2 during the flare and the opposite
sense of circular polarization relative to main flare source 1
confirm that sources 1 and 2 correspond to the footpoints of a flare
loop.

The solid profile in the 4th panel in Figure~\ref{F3--simple} shows
the 34 GHz flux emitted from source 1 and the dashed curve presents
the flux from outside this source. The outside flux was rather small
and the emission was almost entirely produced by source 1. We
therefore conclude that the total power flux recorded with the NoRP
at 35 GHz only contains contribution only from source 1.
 Here we suggest that the emission properties at 34 GHz (NoRH) and 35 GHz (NoRP) are the same.

Although the accuracy of the NoRH image coalignment
 with the magnetogram (about 10 arcsec) is comparable
 with the S--intrusion width we believe that
 the data on the flare configuration allow us
  to associate source 1 with the S--polarity region.
  The value of the photospheric magnetic field at the 34 GHz source site was about 1~kG.
    Data on thermal X--rays were provided by the GOES and low--energy RHESSI channels.
    During the impulsive phase the emission measure was estimated from the
    GOES data
    as $ 5.3\times 10^{48}$ cm$^{-3}$. In the RHESSI data this value is twice as high.
    For an X--ray source size of 10 arcsec we estimate the density
   of the source plasma to be $(6-10)\times 10^{10}$ cm$^{-3}$.

\section{Discussion}

The flare under study demonstrates significant nonthermal particle
signatures with delayed thermal emission.  The flare is initiated by
interaction between a small loop and a large loop. The interaction
site, seen as the main source in hard X--rays and in microwaves, is
located near the common footpoint of the loops. Recently such flares
have been discussed by \cite{Altyntsev16} and \cite{Fleishman16a}.

In our event the spatial observations showed that almost all of the
emission at the NoRH frequencies (17 and 34 GHz) comes from source
1. It is shown that their microwave emission changes the
polarization sense between 17 and 35 GHz. We argued that this source
is located above a narrow S--polarity channel within a large
N--polarity region. The polarization sense reversal above 17 GHz
therefore corresponds to the transition from the extraordinary to
the ordinary wave. Such reversal is opposite to what we expect at
crossing through the spectral peak frequency while the spectrum
transits from the optically thick to the thin regime.

There are two ways to generate the ordinary--mode polarization at
high frequencies in the framework of the gyrosynchrotron emission
mechanism. The first way has recently been proposed by
\cite{Fleishman13}: emission by relativistic positrons. Indeed, if
the flare emission at high frequencies is primarily produced by
high--energy positrons rather than electrons the polarization will
correspond to the ordinary wave mode. The corresponding light curves
recorded with the RHESSI and \textit{Wind}/KONUS spectrometers show
that there was no annihilation line that would be indicative of the
flare positrons during the event under study. The possibility of
producing microwave emission by positrons was first pointed out by
\cite{Lingenfelter}. \cite{Trottet08} found a close correlation in
time and space of radio emission at 210 GHz with the production of
pions during the 2003 October 28 flare. However, an
order--of--magnitude estimate of possible positron numbers
consistent with an X--ray non--detection of the positron
annihilation line rather suggests that the numbers of positrons are
far too small to account for the observed radio emission.  Thus we
have no evidence in support of the positron hypothesis in this
flare.

The other possibility to generate the ordinary--mode emission from
the optically thin source is through pitch--angle anisotropy of the
emitting electrons (\opencite{Fleishman03}). To test this
possibility, we simulated the high--frequency part of the spectra
using a uniform source model with the gyrosynchrotron (GS) fast
codes developed by \cite{Fleishman10}. We found that the
polarization reversal at frequencies above 17~GHz can indeed be
achieved for some angular distributions of emitting electrons.
Specifically, to obtain the polarization reversal as observed we
used the distributions with a Gaussian pitch--angle anisotropy given
by the expression:
$g(\mu)\sim\exp[-(\mu-\mu_{0})^{2})/\Delta\mu^{2}]$, where $\mu_{0}
= \cos\alpha_{0}$ is the beam direction, and $\Delta\mu$ is the beam
angular width.  This expression represents the beam along the field
line for $\mu_{0} = \pm1$, the transverse beam for $\mu_{0} = 0$,
and an oblique beam (or a hollow beam) otherwise.

One of the best 'by eye' fits  is shown by the dashed curve in
Figure~\ref{F7--simple} (panel 3). In this example the polarization
dependence on frequency is similar to the observed dependence at
frequencies above 7 GHz.  The parameters of the best--fit angular
distribution are a source size of $10\times10\times10$ arcsec, a
plasma density and temperature of 20 MK and $3\times 10^{11}$
cm$^{-3}$, a power--law index of $\delta_{MW}$ = 3.5, a density of
the nonthermal electrons of $9\times10^{7}$ cm$^{-3}$, a magnetic
field of 480~G, a viewing angle of $83^{o}$, $\Delta\mu = 0.4$, and
a beam direction $\alpha_{o} = 60^{o}$. The electron index
$\delta_{MW}$ is consistent with the X--ray flux index $\gamma_{X} =
3.0 - 3.2$. Simulations show a strong dependence of the polarization
spectrum on the anisotropy parameter, and this does not match the
constant shape of the polarization spectrum during the burst. In
addition, the density of the background plasma in the model and
X--ray data are mismatches as it is three to five times higher than
the estimates derived from the GOES and RHESSI data. A similar
mismatch was reported by \cite{Fleishman16} for a subclass of
flares, but in these cases the radio and X--ray emissions were
produced by distinct loops. Thus, we conclude that a polarization
reversal through the beam--like anisotropy is unlikely in this
event.

\begin{figure}    

 \centerline{\includegraphics[width=0.8\textwidth,clip=]{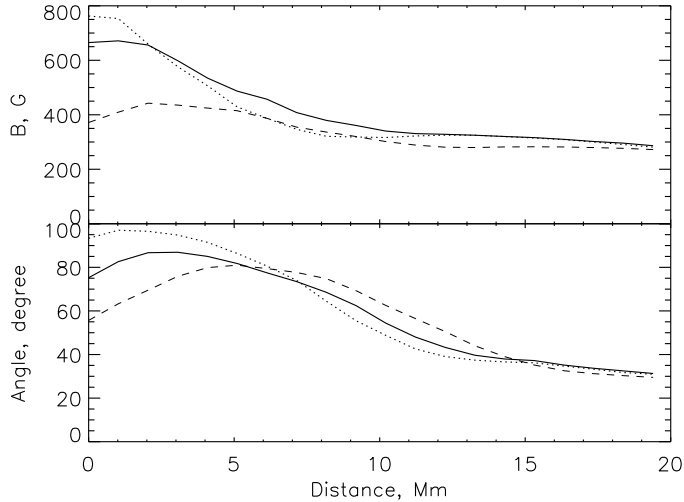}
           }
             \caption{Height dependences of the
             magnetic field magnitude and the angle
             between its direction and the line of
             sight toward source 1 calculated
             for three points. The solid line corresponds
              to the line that starts from the 34 GHz brightness
               center, the dotted line shows the region three arcseconds to the east,
                and the dashed lines show the regions three arcseconds to the south.
                     }
  \label{F8--simple}
  \end{figure}

We consider the possibility that the ordinary wave in the optically
thin mode is observed because it crosses a QT layer on the way from
source 1 to Earth. An unipolar right--handed source at 17 GHz was
observed at the place of source 1 from July 4
(Figure~\ref{QTreversal}). A day before the flare, its sense of
polarization reversed, and this can be regarded as a signature of a
QT layer formation. In this case, the angle between the magnetic
field vector and the line of sight must achieve 90 degrees somewhere
along the propagation path, and the critical frequency $f_{t}$ must
exceed 35~GHz there. The critical frequency is determined by plasma
parameters in the layer $f_{t}\simeq
7.6(N_{10}B^{3}_{100}L_{9})^{0.25}$ GHz, where $N_{10}$ is the
thermal density, expressed in the units of $10^{10}$ cm$^{-3}$;
$B_{100}$ is the magnetic field, expressed in units of 100 G; and
$L_{9}$ is the scale of the angle changing, expressed in units of
$10^9$ {cm} (\opencite{Zheleznyakov}). It has the strongest
dependence on the magnetic field value.

We reconstructed the coronal magnetic field in the nonlinear
force--free approximation, using an SDO/HMI vector magnetogram as
input data. The field reconstruction was performed by the
optimization method, proposed by \cite{Wheatland00}. In the study we
used the implementation of the method developed by \cite{Rudenko09}.
The magnetic field magnitudes calculated along the lines of sight
are shown in Figure~\ref{F8--simple} (top panel). The different
curves represent extrapolations for three points of source 1. The
solid curve corresponds to the line of sight and starts from the 34
GHz brightness center, while the other curves begin at the points
shifted three arcseconds toward east or south. The QT layer most
likely exists at heights below 5 Mm where the magnetic field $B
>$ 500~G. This height is in accordance with the size of the small
loop. The distance  between the footpoints of the small loop was
about 15 arcsec, so the altitude of source 1 might indeed be below
the 5 Mm height. Magnetic field and plasma density in the QT layer
may be sufficient to reverse the polarization height at frequencies
of up to 35 GHz if the layer height is about 5 Mm.

In the case of the QT reversal we fit the microwave spectrum using
an isotropic angular distribution of emitting electrons. One of the
best fits is shown by the solid curve in Figure~\ref{F7--simple}
(panel 3). Here we assumed that $f_{t} > 35$ GHz, and we show the
calculated spectrum in polarization with the opposite sign. The
observed spectrum can clearly be modeled with the reasonable fitting
parameters: a source size is $3\times3\times5$ arcsec, plasma
density $3\times 10^{10}$ cm$^{-3}$, power--law index is
$\delta_{MW}$ = 3.0, density of the nonthermal electrons is
$4.5\times10^{7}$ cm$^{-3}$, magnetic field is 610~G, and viewing
angle is $80^{o}$.

\section{Conclusions}
Spatially resolved microwave polarization observations can provide
important data on flare topology and plasma parameters. We have
studied the flare SOL2012-07-06T01:37 which is a good illustration
of a flare initiated by the interaction of a large and a small loop.
The interaction site is seen as the main flare source in X--rays and
microwaves and is characterized by an unusual behavior of the
polarization spectrum: the wave type changes from extraordinary mode
to ordinary mode at a frequency between 17 and 35 GHz. We have
identified two possible reasons for the observed reversal of
polarization sense. First, it could be due to a pitch--angle
anisotropy of the emitting electrons. The scenario in which the
electrons are strongly beamed places the source above 5 Mm and
requires a dense loop there, which is not confirmed by either X--ray
or EUV data.

On the other hand the spectrum fitting shows that the
 polarization reversal can be explained by the transition of
 gyrosynchrotron emission from the optically thick to
 thin mode. The observed  circular polarizations are
 opposite to the intrinsic polarizations and can be reversed by
  the quasi--transverse effect. The quasi-transverse (QT) scenario places the source
  relatively low in the corona, which releases the
  stringent requirements on the thermal number density
   and nonthermal electrons, given that the magnetic
   field is reasonably large there. Recently, \opencite{Sadykov}
    reported a flare that occurred  in a
system of a low--lying loop arcade with a height of $\le$ 4.5 Mm
using \textit{New Solar Telescope} (NST) data. \cite{Wang} reported
a similar finding for flare precursors using a combination of NST
and EOVSA data. We conclude that flares occurring in rather
low--lying loops may not be unusual.

\section{Acknowledgements}

We thank the anonymous referee for valuable comments. We are
grateful to the teams of the \textit{Siberian Solar Radio Telescope,
Nobeyama Radio Observatory}, \textit{Radio Solar Telescope Network}
(RSTN), and RHESSI, who have provided open access to their data.
This work was supported in part by RFBR grants 15--02--01089,
15--02--03717, 15--02--03835, 15--02--08028, and 16--02--00749, NSF
grants AGS--1250374 and AGS--1262772, and NASA grant NNX14AC87G to
New Jersey Institute of Technology. This study was supported by the
Program of basic research of the RAS Presidium No. 7.

\end{article}

\end{document}